# Effects of Quantum-Well Inversion Asymmetry on Electron-Nuclear Spin Coupling in the Fractional Quantum Hall Regime

Katsushi Hashimoto,[1,2,a] Koji Muraki,[1,b] Norio Kumada,[1] Tadashi Saku,[3] and Yoshiro Hirayama[1,2]

[1]NTT Basic Research Laboratories, NTT Corporation, 3-1 Morinosato-Wakamiya, Atsugi 243-0198, Japan

[2]SORST-JST, 4-1-8 Honmachi, Kwaguchi 331-0012, Japan

[3]NTT Advanced Technology, 3-1 Morinosato-Wakamiya, Atsugi 243-0198, Japan

We examine effects of inversion asymmetry of a GaAs/Al$_{0.3}$Ga$_{0.7}$As quantum well (QW) on electron-nuclear spin coupling in the fractional quantum Hall (QH) regime. Increasing the QW potential asymmetry at a fixed Landau-level filling factor ($\nu$) with gate voltages suppresses the current-induced nuclear spin polarization in the $\nu = 2/3$ Ising QH ferromagnet, while it significantly enhances the nuclear spin relaxation at general $\nu$. These findings suggest that mixing of different spin states due to the Rashba spin-orbit interaction strongly affects the electron-nuclear spin coupling.

---

[a] Email address: khashimo@physnet.uni-hamburg.de. Present address: Institute of Applied Physics, Hamburg University, Jungiusstraβe 11, D-20355 Hamburg, Germany.

[b] Email address: muraki@will.brl.ntt.co.jp.

A search for electrical means to access spins in the solid state, particularly in semiconductors, has recently been accelerated by interests in basic physics as well as technological requirements from the emerging fields of quantum information [1] and spintronics [2]. Various semiconductor devices using both electronic [3] and nuclear [4] spins have been proposed. For the case of electron spin, one key ingredient is the spin-orbit interaction. In semiconductor heterostructures, the spin-orbit Hamiltonian has two relevant contributions: the Dresselhaus term arising from the bulk inversion asymmetry of the host material and the Rashba term arising from the structural inversion asymmetry of the heterostructure [5]. Of particular interest is the latter, which can be controlled by a gate bias as experimentally demonstrated for narrow-gap semiconductors such as $In_xGa_{1-x}As$ ($x \geq 0.5$) [6,7] that inherently have strong Rashba coupling.

On the other hand, electrical access to nuclear spins requires mediation through the hyperfine interaction. This underlies the renewed interests in the two dimensional electron system (2DES) under high magnetic fields, namely, the quantum Hall (QH) system, where various nuclear-spin-related phenomena have been reported [8,9,10,11,12,13,14,15]. At Landau-level filling factor ($\nu$) of 2/3, when spin-polarized and unpolarized ground states become energetically degenerate at the phase transition point, the system becomes an Ising QH ferromagnet [16], and current flowing across ferromagnetic domain boundaries causes flip-flop scattering of electrons with nuclear spins, which dynamically polarizes nuclei and causes the longitudinal resistance ($R_{xx}$) to grow over a long time scale [10,11,12,13]. In contrast, nuclear spin relaxation is drastically enhanced at filling factors slightly deviated from $\nu = 1$ through the coupling with low-energy spin modes that accompany the non-colinear spin structure, i.e., Skyrmions, of the 2DES [17]. Previous studies have exploited these to demonstrate electrical control of nuclear spin polarization and relaxation via tuning the filling factor [11,15].

In this Letter, we report that the coupling of electronic and nuclear spins in the fractional QH regime is significantly altered when a strong electric field is applied perpendicular to the 2DES at a fixed filling factor. Using a 20-nm-wide GaAs/Al$_{0.3}$Ga$_{0.7}$As quantum well (QW) with both front and back gates, we show that the nuclear-spin-related $R_{xx}$ anomaly in the $\nu = 2/3$ Ising QH ferromagnet is completely suppressed when the QW is made strongly asymmetric. We separately demonstrate that the potential asymmetry also affects the nuclear spin relaxation rate at general $\nu$. The result that the potential asymmetry plays an essential role in both experiments suggests that the Rashba effect, generally believed to be irrelevant in GaAs [18], can become important in tailoring the coupling between electronic and nuclear spins. We discuss possible mechanisms by which the spin-orbit interaction affects the electron-nuclear spin coupling.

The 20-nm-wide GaAs/Al$_{0.3}$Ga$_{0.7}$As QW was grown following an AlAs/GaAs superlattice barrier on a Si-doped $n^+$-GaAs (100) substrate that is used as a back gate. The QW is modulation-doped on the front side, and has an as-grown electron density of $n = 1.7 \times 10^{15}$ m$^{-2}$. The sample is processed into a 50-$\mu$m-wide Hall bar with voltage probes at an interval of 180 $\mu$m. By adjusting the voltages of both front and back gates, we tune the electron density and the QW potential asymmetry independently. As a measure of the potential asymmetry, we take the difference, $\delta n = n_b - n_f$, between the densities of electrons supplied from the back side ($n_b$) and front side ($n_f$) of the QW. The QW is then expected to be symmetric at $\delta n = 0$ m$^{-2}$. In our device, $\delta n$ can be varied from $-2.1 \times 10^{15}$ to $4.7 \times 10^{15}$ m$^{-2}$. The mobility at $n = 1.3 \times 10^{15}$ m$^{-2}$ is 130 m$^2$V$^{-1}$s$^{-1}$ for $\delta n = 0$ m$^{-2}$, while it slightly decreases to 120 m$^2$V$^{-1}$s$^{-1}$ for $\delta n = 4.7 \times 10^{15}$ m$^{-2}$. Unless otherwise specified, measurements of $R_{xx}$ were carried out at 130 mK, using a standard lock-in technique and a drain-source current $I_{ds} = 0.5$ nA at a *normal* sweep rate, $d\nu/dt \sim 0.2$ min$^{-1}$, for which effects of nuclear spin polarization are negligible. For time-dependent measurements, nuclear spins were randomized prior to each run by temporarily setting $\nu$ to around 0.9 or 1.1 for 200 s [11].

Before addressing the effects of the potential asymmetry on the electron-nuclear spin coupling, we first describe how the potential asymmetry affects the magnetic field position of the $\nu = 2/3$ phase transition. Figure 1(a) shows two color-scale plots of $R_{xx}$ as functions of $\nu$ and $\delta n$ at magnetic fields of $B = 7.64$ and 8.27 T. The $\nu = 2/3$ QH region is separated into spin-polarized and unpolarized phases located at lower and higher $\nu$ [13]. At $B = 7.64$ T, the transition point, manifested by the $R_{xx}$ peak separating the two regions, is located exactly at $\nu = 2/3$ near $\delta n = 0$ m$^{-2}$. Upon increasing $|\delta n|$, however, the transition point shifts toward smaller $\nu$ as a result of changes in the g-factor [19] and Coulomb interactions due to the increased confinement. To avoid possible influence of this deviation of $\nu$ away from 2/3, we adjusted the magnetic field for each $\delta n$ such that the transition occurs just at $\nu = 2/3$, e.g., by setting $B = 8.27$ T for $\delta n = 4.7 \times 10^{15}$ m$^{-2}$ [Fig. 1(a), bottom]. The magnetic field, $B_{t,\,\nu=2/3}$, thus chosen for each $\delta n$ traces a parabola centered at $\delta n \approx 0$ m$^{-2}$ [Fig. 1(b)]. This confirms that the potential is symmetric at $\delta n \approx 0$ m$^{-2}$, and becomes asymmetric with increasing $|\delta n|$.

We now examine effects of the potential asymmetry on the electron-nuclear spin coupling. At $\delta n = 0$ m$^{-2}$, when the filling factor is swept at an extremely *slow* rate (d$\nu$/d$t$ ~ 7 × 10$^{-4}$ min$^{-1}$) with high current ($I_{ds}$ = 10 nA), the $R_{xx}$ peak at the phase transition point is dramatically enhanced over that taken by the *normal* scan [Fig. 1(c), top]. This is due to dynamic nuclear spin polarization induced by flip-flop scattering upon transport across boundaries between spin-polarized and unpolarized domains [10]; it gives rise to spatially inhomogeneous hyperfine fields and hence additional source of scattering [12]. Now we make the QW strongly asymmetric ($\delta n$ = 4.7 × 10$^{15}$ m$^{-2}$) [Fig. 1(c), bottom]. We find that the $R_{xx}$ enhancement, and hence the nuclear spin polarization, is totally suppressed; the two curves for the normal and slow scans completely overlap. Such a striking effect is unlikely to be caused by either the slight decrease in the mobility (4%) or the increase in the density (8%). This is evident from the fact that a larger $R_{xx}$ enhancement has been reported for samples with a higher electron density (11% higher than in our sample) [10] or lower mobility [14].

To prove that the observed effect is indeed associated with the potential asymmetry, we have studied the dependence of the $R_{xx}$ enhancement on $\delta n$ in more detail. Figure 2 (a) presents time dependence of $R_{xx}$ for different values of $\delta n$ taken at the $\nu$ = 2/3 transition point with $I_{ds}$ = 10 nA. When the QW is almost symmetric ($\delta n$ = 0.18 × 10$^{15}$ m$^{-2}$), $R_{xx}$ increases with time and saturates after about 8 min. As $\delta n$ is increased, the $R_{xx}$ enhancement becomes smaller until it is no longer discernible by $\delta n = 4.12 \times 10^{15}$ m$^{-2}$. In Fig. 2 (b), we plot $\Delta R_{xx}$, the increase in $R_{xx}$ from its initial value, as a function of $\delta n$. The data demonstrate that $\Delta R_{xx}$ is largest near $\delta n = 0$ m$^{-2}$, and decreases with increasing $|\delta n|$ [20]. This clearly shows that the potential asymmetry plays an essential role. While the $R_{xx}$ enhancement is known to strongly depend on $I_{ds}$ [10,12], we confirm similar $\delta n$ dependence up to $I_{ds} = 20$ nA. [Above 20 nA (not shown), strong electron heating effects are observed.]

The observed effect of the potential asymmetry suggests involvement of the Rashba spin-orbit interaction. Indeed, there are theories describing the effects of spin-orbit interaction on Landau-level coincidences [21] and associated QH ferromagnetisms at integer fillings [22] [23]. Within a single-particle picture, the spin-orbit interactions mix Landau levels with different spins, thereby creating a small anticrossing gap [21]. For the present case of an Ising QH ferromagnet, where dissipative transport occurs via excitation modes within domain boundaries [24], the role of the spin-orbit interaction is to induce a finite gap to an otherwise gapless mode [23,25]. We have therefore carried out activation studies at the $\nu = 2/3$ transition point for different $\delta n$ [Fig. 3(a)], which reveal that the energy gap ($\Delta$), deduced via $R_{xx} \propto \exp(-\Delta/2k_B T)$, takes a minimum near $\delta n = 0$ m$^{-2}$ and does increase with $|\delta n|$ [Fig. 3(b)]. We note that even at $\delta n = 0$ m$^{-2}$, $\Delta$ is finite at 8 $\mu$eV, which may originate from Coulomb interactions [24] and/or the Dresselhaus-type spin-orbit interaction. For the largest asymmetry of $\delta n = 4.7 \times 10^{15}$ m$^{-2}$, $\Delta$ increases by 19 $\mu$eV, i.e., by 240% of its value at $\delta n = 0$ m$^{-2}$. Although this may include the change in the Coulomb energy $E_c$, here $E_c$ $\left(\propto \sqrt{B}\right)$ increases by only 4%, which alone cannot account for this large effect. It is thus most likely that the Rashba spin-orbit interaction, which is absent for $\delta n = 0$ m$^{-2}$, becomes operative for $\delta n \neq 0$ m$^{-2}$ and increases $\Delta$ [26][27].

Now we discuss the mechanism by which the spin-orbit interaction affects the electron-nuclear spin coupling. We first recall that the current-induced nuclear spin polarization occurs as a result of angular momentum conservation, which requires that a flip of an electron spin upon transport across domain boundaries must be accompanied by a simultaneous flop of a nuclear spin in the opposite direction. On the other hand, the spin-orbit interaction mixes Landau levels with different spins, which provides an alternative path for electrons to flip their spins. Indeed, the spin-orbit interaction is believed to be the dominant mechanism for the spin-flip scattering between spin-split integer QH edges [28]. Hence, a strong spin-orbit coupling would allow for transport across domain boundaries without nuclear spin flops, resulting in the reduced spin transfer from the electronic system to the nuclear system. Note that the larger $\Delta$ implies a higher transmission probability across domain boundaries [23], supporting this picture.

Besides the spin transfer efficiency discussed above, the nuclear spin relaxation rate ($T_1^{-1}$) is also expected to contribute to the reduced polarization. We therefore have examined how the potential asymmetry affects $T_1^{-1}$. This was done at a fixed magnetic field of 7.6 T in the following sequence [Fig. 4(a)]. First, nuclear spin polarization is generated at the $\nu = 2/3$ transition point with $I_{ds} = 20$ nA and $\delta n = 0$ m$^{-2}$ until $R_{xx}$ saturates. We measure the change in $R_{xx}$ after a given period of time $\tau$, during which $\delta n$ as well as $\nu$ are set to temporal values, $\delta n_\text{temp}$ and $\nu_\text{temp}$. The relaxation of $R_{xx}$ after $\tau$ thus represents the degree of nuclear spin relaxation for the condition $(\nu, \delta n) = (\nu_{\text{temp}}, \delta n_{\text{temp}})$. Figure 4(a) shows data for $\nu_{\text{temp}} = 0.5$ and $\tau = 5$ s, where cases for $\delta n_{\text{temp}} = 0$ and $4 \times 10^{15}$ m$^{-2}$ are compared. It is seen that the relaxation of $R_{xx}$ is larger for $\delta n_{\text{temp}} = 4 \times 10^{15}$ m$^{-2}$. This clearly indicates that the nuclear spin relaxation rate is not a unique function of $\nu$, but also affected by $\delta n$.

Figure 4(b) compiles similar measurements carried out for $\nu_{\text{temp}} = 0.3$-$1.25$ and $\tau = 2$ s, where $R_{xx}$ just after the relaxation procedure (symbols) is compared with the saturation value before the relaxation (dashed line). The data reveal that the enhancement of the nuclear spin relaxation with $\delta n$ is ubiquitous; it occurs for general $\nu$ over a wide range[29]. We have performed measurements for different values of $\tau$ to extract $T_1^{-1}$. Results for $\nu_{temp} = 0.5$ and $0.75$ are shown in Fig. 4(c) as a function of $\delta n_{\text{temp}}$. The data clearly show that $T_1^{-1}$ is enhanced with $|\delta n|$, again demonstrating the importance of the potential asymmetry. We note that even for the largest $|\delta n|$ the peak amplitude of the electron probability distribution increases by only 23%, which cannot account for the large enhancement of $T_1^{-1}$ observed.

The enhanced $T_1^{-1}$ may suggest that the electron-nuclear spin coupling is strengthened, seemingly inconsistent with the suppression of the current-induced nuclear spin polarization at $\nu = 2/3$. We however suggest below a possible mechanism by which the spin-orbit coupling can enhance $T_1^{-1}$. As already discussed, the spin-orbit interaction mixes Landau-levels with different spins, making the electron spin component along the magnetic field direction no longer a conserved quantity even in the presence of a strong

Zeeman coupling. This will cause local spin direction of electrons to vary with time and therefore yields fluctuating hyperfine fields. If such fluctuations occur at sufficiently low frequencies such that the nuclei can follow, the nuclear spins will be randomized, leading to enhanced $T_1^{-1}$ [30]. Since the relaxation measurement presented here is applicable only to filling factors other than $\nu = 2/3$, the relative importance of the two mechanisms in the $\nu = 2/3$ QH ferromagnet is not known. Nevertheless, the two striking effects due to the potential asymmetry demonstrate the importance of the spin-orbit interaction in electron-nuclear spin coupling. Solid understanding of the data, however, must await theories clarifying the roles of spin-orbit interaction in the fractional QH regime.

In summary, we have demonstrated that the coupling of electronic and nuclear spins in the fractional QH regime is significantly modified by the potential asymmetry of the QW. We suggest that the Rashba spin-orbit coupling facilitates electron spin flips, which reduces the net spin transfer to the nuclei and, on the other hand, enhances the nuclear spin relaxation by creating fluctuating hyperfine fields.

The authors are grateful to J. Nitta, Y. Tokura, K. von Klitzing, A. H. MacDonald, J. Schliemann, and K. Takashina for fruitful discussions.

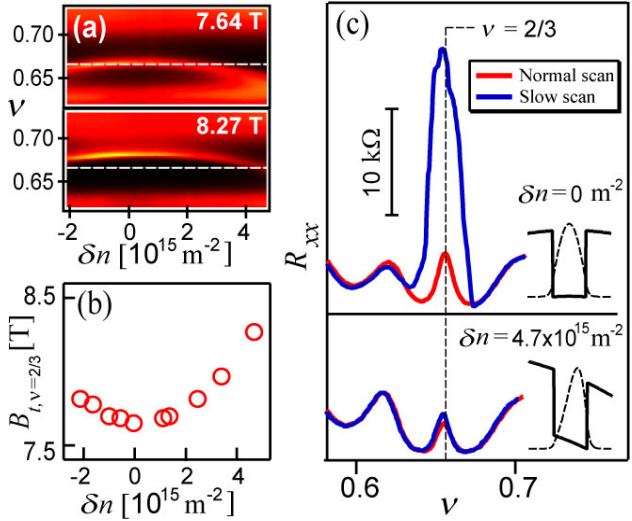

Fig. 1: (a) Color-scale plots of $R_{xx}$ around $\nu = 2/3$ as functions of $\nu$ and the potential asymmetry $\delta n$ at fixed magnetic fields. Dark regions represent small values of $R_{xx}$. Dashed lines indicate $\nu = 2/3$. ($\delta n$ is evaluated at $\nu = 2/3$.) Data are taken by sweeping $\nu$ at a normal rate ($d\nu/dt \sim 0.2$ min$^{-1}$) with low current ($I_{ds} = 0.5$ nA). The magnetic field, $B_{t,\nu=2/3}$, at which the phase transition occurs at $\nu = 2/3$ is determined for each $\delta n$ and plotted in (b). (c) Effects of potential asymmetry on the slow $R_{xx}$ enhancements at $T = 80$ mK. Red and blue curves are taken at normal and slow ($d\nu/dt \sim 7 \times 10^{-4}$ min$^{-1}$) scans, respectively, with high current ($I_{ds} = 10$ nA). Insets: self-consistent potential profiles and wave functions.

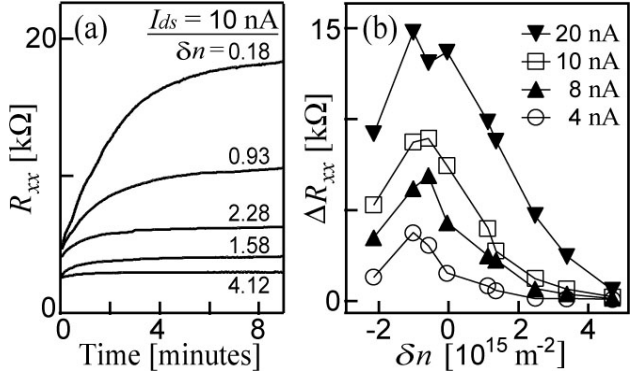

**Fig. 2: (a) Time dependence of $R_{xx}$ at the $\nu = 2/3$ transition point with $I_{ds}$ = 10 nA for different values of $\delta n$ (indicated in the figure in units of $10^{15}$ m$^{-2}$). $T$ = 80 mK. (b) Value of the $R_{xx}$ enhancement versus $\delta n$ for different values of $I_{ds}$.**

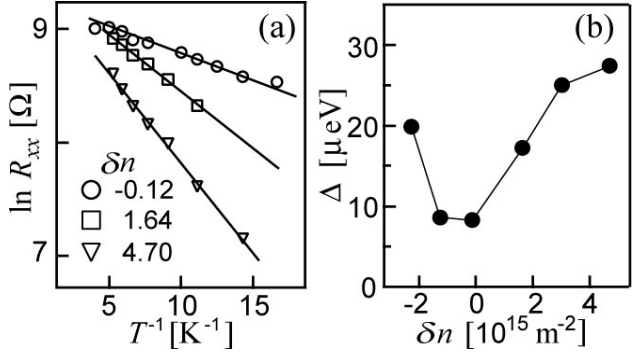

**Fig. 3: (a) Arrhenius plots of the $R_{xx}$ peak at the $\nu = 2/3$ transition point for different $\delta n$ (indicated in the figure in units of $10^{15}$ m$^{-2}$). Solid lines represent fitting with $R_{xx} \propto \exp(-\Delta/2k_B T)$. (b) Energy gap $\Delta$ as a function of $\delta n$.**

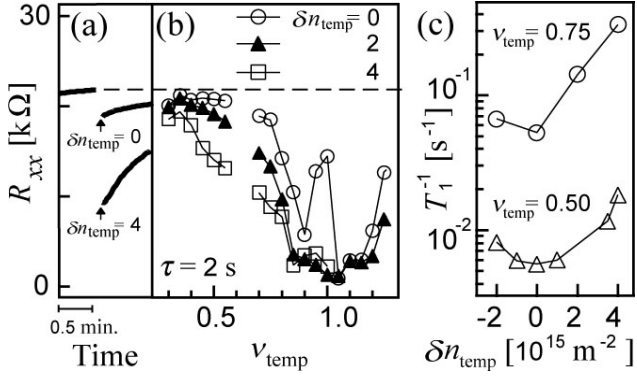

**Fig. 4:** Relaxation measurements at $B = 7.6$ T. (a) Variation of $R_{xx}$ at the $\nu = 2/3$ transition point due to the relaxation procedure of temporarily setting $\nu_{temp} = 0.5$ with $\delta n_{temp} = 0$ and $4 \times 10^{15}$ m$^{-2}$. $I_{ds} = 20$ nA. (See texts for details.) (b) $R_{xx}$ at the $\nu = 2/3$ transition point measured just after the relaxation procedure as a function of $\nu_{temp}$. Dashed line is the saturation value before the relaxation. (c) Relaxation rate $T_1^{-1}$ for $\nu_{temp} = 0.5$ and $0.75$ as a function of $\delta n_{temp}$.

primarily from the lowest Landau level, are not straightforward. Landau-level mixing due to Coulomb interactions and/or disorder must therefore be involved.

[26] Self-consistent calculations show that the average electric field (excluding the QW potential) is $3.3 \times 10^6$ Vm$^{-1}$ for $\delta n = 4.7 \times 10^{15}$ m$^{-2}$. We calculate the Rashba parameter $\alpha$ for this $\delta n$ to be 2.5 (1.5) $\times 10^{-13}$ eVm with (without) the interface effects [7]. (Parameters taken from L. Wissinger *et al.*, Phys. Rev. B **58**, 15375 (1998).] Note that these values are comparable to the Dresselhaus parameter $\beta = 2.6 \times 10^{-13}$ eVm $\left(= \gamma \langle k_z^2 \rangle\right)$ calculated for $\delta n = 0$ m$^{-2}$. [$\gamma = 16.5$ eVÅ$^3$ taken from B. Jusserand *et al.*, Phys. Rev. B **51**, 4707 (1995).]

[27] Though $\beta \left(\propto \langle k_z^2 \rangle\right)$ also increases with $|\delta n|$ (by 40% at maximum), it alone is not enough to account for the large changes in $\Delta$.

[28] A. V. Khaetskii, Phys. Rev. B **45**, 13777 (1992); G. Müller *et al.*, Phys. Rev. B **45**, 3932 (1992).

[29] The fast relaxation near $\nu_{temp} = 1$ is due to the low-energy spin modes associated with Skyrmions. We here focus on the changes due to $\delta n_{temp}$.

[30] We note that a similar argument has been used for $T_1^{-1}$ in quantum dots. [Y. B. Lyanda-Geller, I. L. Aleiner, and B. L. Altshuler, Phys. Rev. Lett. **89**, 107602 (2002).]